\documentstyle[aps]{revtex}

\def\lta{~\raise.4ex\hbox{$<$}\llap{\lower.6ex\hbox{$\sim$}}~}
\def\gta{~\raise.4ex\hbox{$>$}\llap{\lower.6ex\hbox{$\sim$}}~}

\begin{document} \input psfig.sty  

\title{Exploring the cooperative regimes in a model of agents 
without memory or "tags": indirect reciprocity vs. selfish incentives}
\author{H. Fort}

\address{Instituto de F\'{\i}sica, Facultad de Ciencias, Universidad de la
Rep\'ublica, Igu\'a 4225, 11400 Montevideo, Uruguay}

\maketitle

\begin{abstract}

The self-organization in cooperative regimes in a simple mean-field
version of a model based on "selfish" agents which play 
the Prisoner's Dilemma (PD) game is studied. 
The agents have no memory and use strategies 
not based on direct reciprocity nor 'tags'. Two variables are 
assigned to each agent $i$ at time $t$, 
measuring its capital $C(i;t)$ and its 
probability of cooperation $p(i;t)$. At each time step $t$ a pair of agents
interact by playing the PD game. 
These 2 agents update their probability of cooperation $p(i)$
as follows: they compare the profits they made in this interaction 
$\delta C(i;t)$ with an estimator $\epsilon(i;t)$
and, if $\delta C(i;t) \ge \epsilon(i;t)$, agent $i$ 
increases its $p(i;t)$ while if $\delta C(i;t) < \epsilon(i;t)$ the agent 
decreases $p(i;t)$. 
The 4!=24 different cases produced by permuting the four
Prisoner's Dilemma canonical payoffs 3, 0, 1, and 5 - corresponding,
respectively, to $R$ (reward), $S$ (sucker's payoff), $T$ 
(temptation to defect) and $P$ (punishment) - are analyzed.
It turns out that for all these 24 possibilities, after a transient,   
the system self-organizes into a stationary state with average equilibrium
probability of cooperation $\bar{p}_\infty$ = constant $ > 0$.
Depending on the payoff matrix, there are different equilibrium 
states characterized by their average probability of cooperation
and average equilibrium per-capita-income 
($\bar{p}_\infty,\bar{\delta C}_\infty$).

\end{abstract}

\vspace{2mm}

{\it keybords}: Complex adaptive systems, Agent-based models, Social systems

\vspace{1mm}


\vspace{2mm}

\section{Introduction}

A common approach to the problem of how cooperation emerges in 
societies of "selfish" individuals - individuals which pursue exclusively 
their own self-benefit - is based on game theory, and specifically 
on the {\it Prisoner's Dilemma} (PD) of the early fifties.
In a series of works Robert Axelrod and co-workers 
\cite{axel84} 
used this kind of computer games to examine the basis of cooperation 
between selfish agents in a wide variety of contexts. 
Mechanisms of cooperation based on the PD have shown their 
usefulness in economy \cite{ff02}-\cite{w89}, 
political science \cite{h90}-\cite{g88}, international relations 
theory \cite{h01}-\cite{s71}, 
theoretical biology \cite{wn99}-\cite{n90}, ecosystems \cite{md97}-
\cite{dmh92}, etc. 

The beauty of the PD game relies on the fact that it embodies 
the central ingredients of the cooperation problem in a very simple 
and intuitive way. There are two players, each confronting two choices:
cooperate (C) or defect (D) and each makes its choice without
knowing what the other will do. Independently of what the other 
player does, defection D yields a higher payoff than cooperation
and is the dominant strategy.
In other words, the outcome (D,D) of both players is the Nash 
equilibrium \cite{n51}.
The dilemma is that if both defect, both do worse than if
both had cooperated.

The emergence of cooperation in prisoner's dilemma (PD) games is generally 
assumed to require repeated play (and strategies such as 
Tit for Tat (TFT) \cite{axel84}, 
involving memory of previous interactions) or features ("tags") 
permitting cooperators and defectors to distinguish one another 
\cite{ep98}.

In this work, 
I consider a simple model of selfish agents playing PD, possessing 
neither memory nor tags, to study the self-organized cooperative states 
which emerge for different payoff matrices.
The model consists of  $N_{ag}$ agents,  with
two variables assigned to each agent at the site or cell $i$ and
at time $t$: its probability of cooperation $p(i;t)$ and its capital 
$C(i;t)$. Pairs of agents, 1 and 2, interact by playing the PD game at 
each time step $t$.
That is, there are 4 possible outcomes for the interaction of agent $i$ 
with agent $j$ : 1) they can both 
cooperate (C,C) 2) both defect (D,D), 3) $i$ cooperates and $j$ defects
(C,D) and 4) $i$ defects and $j$ cooperates (D,C). 
Depending on the situation 1)-4), the agent $i$ ($j$) gets respectively
: the "reward" $R (R)$, the "punishment" $P (P)$, the "sucker's 
payoff"$ S (the "temptation to defect" T)$ or $T (S)$, 
{\it i.e.} the payoff matrix M$^{RSTP}$ is

\begin{center}

$${\mbox M}^{RSTP}=\left(\matrix{(R,R)&(S,T)\cr (T,S)&(P,P) \cr}\right)$$

\end{center}

The payoff matrix gives the payoffs for ROW actions when confronting
with COLUMN actions.

After playing the PD the agents update their probability of cooperation 
$p(i;t)$ and $p(j;t)$ according to the same definite "measure of success" 
which does not vary with time . Thus all agents follow a universal and 
invariant strategy defined by a measure of success plus an updating 
rule to transform $p(1;t)$ and $p(2;t)$ into $p(1;t+1)$ and $p(2;t+1)$. 

The 4!=24 different payoff matrices produced by permutation of the four
Prisoner's Dilemma canonical payoffs -3, 0, 1, and 5- are analyzed
by means of a Mean Field (MF) approach, in which all the spatial correlations 
in the system are neglected.
It turns out that for all these 24 possibilities, after a transient,   
the system self-organizes into a state of equilibrium
characterized by the average probability of cooperation and
average per-capita-income ($\bar{p}_\infty,\bar{\delta C}_\infty$), 
always with $\bar{p}_\infty > 0$. Furthermore, in the majority
of cases $\bar{p}_\infty \gta 0.5$.

Payoff matrices can be classified into sub-categories 
according to their dominant strategy. 
Let us call $M_D$ the class of those matrices such that:
\begin{equation}
T>R, \;\; \mbox{and} \;\; P>S,
\label{eq:class1}
\end{equation}
for which the dominant strategy is D. This class comprises six
matrices: M$^{3051}$, M$^{1053}$, M$^{1035}$, M$^{0315}$, M$^{0153}$
and M$^{0135}$.
A second class $M_C$ corresponds to 
\begin{equation}
R>T, \;\; \mbox{and} \;\; S>P,
\label{eq:class2}
\end{equation}
for which the dominant strategy is C and comprises the
following six matrices: M$^{5310}$, M$^{5301}$, M$^{5130}$, M$^{3510}$, 
M$^{3501}$ and M$^{1503}$.
The remaining twelve matrices do not comply with
equation (\ref{eq:class1}) or (\ref{eq:class2}) and produce situations
not dominated by (D,D) or (C,C).

The only payoff matrix that implies a dilemma, in the sense 
explained above, is the 
canonical one with $R=3, S=0, T=5$ and $P=1$ which belongs to class
$M_D$ and comply 
with the condition (\ref{eq:class1}) plus condition $R>P$, 
or equivalently the chain of inequalities: $T>R>P>S$ 
\footnote{In fact there is an "anti-dilemma"
posed by matrix M$^{1503}$ for which $T<R<P<S$
and although the dominant strategy is C
both players would prefer the punishment P associated with (D,D).}.
However, some matrices exhibit a tension between C and D 
and give rise to $\bar{p}\simeq \frac{1}{2}$. 
The matrices which do not embodie such 
trade-off produce the situations which depart from 
$\bar{p}_\infty \simeq \frac{1}{2}$. Clearly, these payoff matrices 
are unrealistic in order to model the social behavior of the majority
of individuals.
So, why bother to study matrices which imply no dilemma?
Well, one reason is that they could be of importance in other contexts. 
One might envisage situations in which a definite value of 
$\bar{p}_\infty$ is required in the design
of a system or is the one which optimizes the functioning of a 
particular mechanism, etc.
Another motivation is that this "unreasonable" payoff matrices 
can be used by minorities of individuals which depart from the "normal" ones 
(assumed to be neutral) for instance, D-inclined "free riders" or 
C-inclined "altruistic" individuals.
Finally, we will show results for these payoff matrices which, at first
glance, defy our intuition. For example, payoff matrices which, at least in
principle, one would bet that favor cooperation and indeed produce a very 
low degree of cooperation.

\section{The Model}

The pairs of interacting partners, by virtue of the MF treatment,
are chosen randomly instead of being restricted to some 
neighborhood. The implicit assumptions are that the population is
sufficiently large and the system connectivity is high {\it i.e.} the
agents display high mobility or they experiment interaction at a 
distance (for instance electronic transactions). 
In this work the population of agents will be fixed to $N_{ag}=1000$
and the number of time steps will be of order $t_f=10^5-10^6$ in such
a way that both assumptions be also consistent with the fact that 
agents have no memory.

Starting from an initial state at $t=0$ taken as $p(i;0)$ chosen at random 
( in the interval [0,1] ) and $C(i,0)$ = 0 for each cell $i$, 
the system evolves by iteration during $t_f$ time steps following these
stages in this order:

\begin{enumerate}

\item{\it Selection of players:} 
At each time step $t$ two agents, located at random positions $i$ 
and $j$, are selected to interact {\it i.e.} for playing the PD game. 
 
\item{\it Playing pairwise PD:}  
The action, C or D, of each interacting agent $k$
( $k$=$i$ or $k$=$j$ ) is decided generating a random number $r$
and if $p(k;t) > r$ then it cooperates and, conversely, if
$p(k;t) < r$ it defects.

\item{\it Capital update:}
As a result of the interaction the capital of each interacting 
agent $k$ is updated as $C(k;t) \longrightarrow 
C(k;t)+\delta C(k;t)$, being the profit of agent $k$, $\delta C(k;t)$ 
one of the four PD payoffs: $R$, $S$, $T$ or $P$.

\item{\it Assessment of success:}
Each of the two agent who have just interacted 
compares its profit $\delta C(k;t)$
with an {\em estimate} $\epsilon(k;t)$ of the expected utilities.
If $\delta C(k;t) \ge \epsilon(k;t)$ ($\delta C(k;t) < \epsilon (k;t)$ ) 
the agent assumes it is doing well (badly)
and therefore its level of cooperation is adequate (inadequate).

\item{\it Probability of cooperation update:}
Pursuing to increse their utilities in future PD games the agents that 
just interacted update their $p(k;t)$. 
If agent $k$ is doing well it increases its 
probability of cooperation $p(k;t)$ choosing an uniformly distributed
value between $p(k;t)$ and 1. On the other hand,  
if agent $k$ is doing badly it decreases its 
probability of cooperation $p(k;t)$ choosing an uniformly distributed
value between 0 and $p(k;t)$  (see below for 
a discusion of this update rule).

\end{enumerate}

\vspace{2mm}

Let us see how the estimate $\epsilon(k;t)$ emerges naturally.
If the interacting agents $i$ and $j$  
cooperate with probabilities $p_i$ and $p_j$ respectively 
(and defect with probabilities $1-p_i$ and $1-p_j$ ), then 
the expected value of the payoff to $i$, $\delta C^{RSTP}_i$ is given by:
\begin{equation}
\delta C^{RSTP}_i  =  R p_i p_j + S p_i (1-p_j) + T (1-p_i) p_j
+ P (1-p_i) (1-p_j).
\label{eq:Deltacap}
\end{equation}
Hence I consider an estimate $\epsilon(k;t)$, which only
involves the probability of cooperation $p(k;t)$ of the agent $k$ who
uses the estimate, obtained by
replacing in equation (\ref{eq:Deltacap}) $p_i$ and $p_j$ by $p(k;t)$:
\begin{equation}
\epsilon^{RSTP}(k;t) =  (R-S-T+P) p(k;t)^2 +(S+T-2P) p(k;t) +P.
\label{eq:LocalEst}
\end{equation}

While the measure of success seems  
natural, the updating rule for the probability of 
cooperation is quite arbitrary. For instance, for the case of the 
canonical payoff matrix, the
update rule for the probability of cooperation implies the following: if
your partner cooperated, increase your level of cooperation; else lower it
(of course, with boundaries at 0 and 1). 
A priori, it is not obvious if this is a good update rule in order 
to maximize your utilities. After all, the other player might be a 
sucker. In that case perhaps you should defect more. However, as we will see
in the next section, this update rule basically works by tuning
the agent's cooperation in order to accomplish some sort of 
"indirect reciprocity" which in turn produces cooperative equilibrium states.  
Furthermore, modifying the strategy 
of each agent so that it defects more often, when it is doing well, 
pursuing to exploit an assumed high percentage of suckers, ends by
spoiling cooperation.
A natural implementation of this change of strategy would be: if you are 
doing well decrease your probability of cooperation $p$ with probability 
proportional to $1-p$ and increase it with probability proportional to $p$.
But this is equivalent to the {\em replicator dynamics} 
\cite{tj78} - \cite{hs88} for which, in ordinary situations, it is known 
that the cooperation becomes extinct.

\vspace{2mm}

{\it General remarks on the model.}

\vspace{1mm}

{\it I.} Among the weaknesses of major approaches that have been 
considered to answer the question about the emergence of cooperation
two are often remarked. 

{\it I.A} The first criticism is about the generally 
assumed "binary" 
probability of cooperation {\it i.e.} agents either always cooperate (C) or 
always defect (D). Clearly, this is no very realistic. Indeed the levels of
cooperation of the individuals are continuous. 
Hence a real $p(i;t)$ (in the interval [0,1]) is used, reflecting 
existence of a "gray scale" of levels of cooperation instead of just
"black" and "white".

{\it I.B} The second objection
is concerning the deterministic nature of the algorithms which seem to fail
to incorporate the stochasticity of agent behavior. 
The used algorithm is non deterministic. Comparison with the random 
number $r$ reflects a stochastic component of agents behavior.

\vspace{1mm}

{\it II.} All the agents follow the same 
universal strategy which does not evolve over time. However, 
the system is adaptive in the sense that the
probabilities of cooperation of the agents do evolve. 

\vspace{1mm}

\section{RESULTS}

For all the 24 payoff matrices the system self-organizes, after a transient,
in equilibrium states with six values of $\bar{p}_\infty > 0$:
1, 0.56$\pm 0.003$, 0.5$\pm 0.02$, 0.42$\pm 0.006$,
0.22$\pm 0.002$ and 0.115$\pm 0.005$. The 24 measures are performed 
over 100 simulations of $10^6$ time steps each.
Fig. 1 show the average probability of cooperation for different payoff
matrices vs. time for the 200,000 first time steps. 

Roughly, the equilibrium asymptotic states can be classified in 
3 classes: {\em highly cooperative} ($\bar{p}_\infty>0.5$), 
{\em moderately cooperative} ($\bar{p}_\infty \simeq 0.5$) and
of {\em loow cooperation} ( $\bar{p}_\infty < 0.5$).
In the second column of Table 1 are listed the values of 
$\bar{p}_\infty$ for the 24 payoff matrices. 
\begin{center}
\begin{figure}[h]
\centering
\psfig{figure=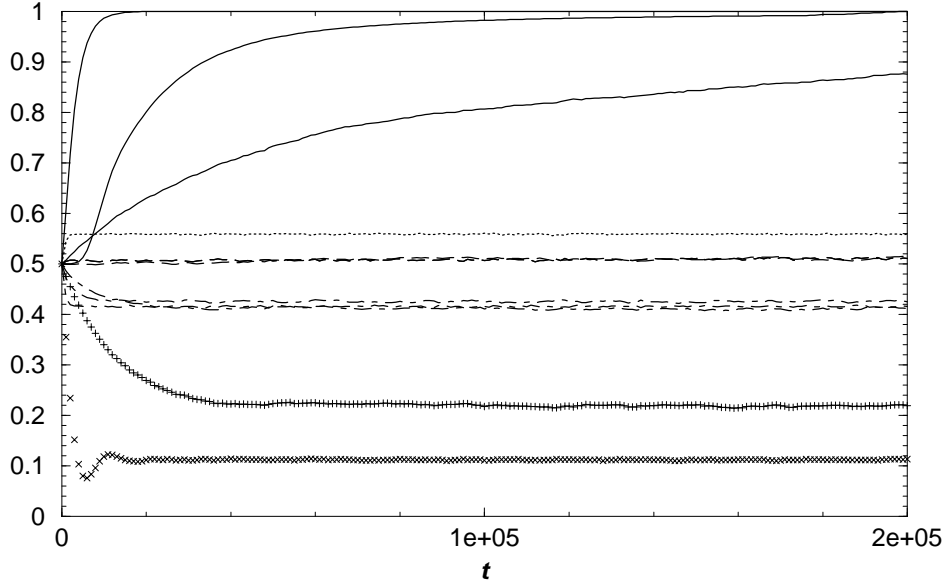,height=8cm}
\caption{Curves of $\bar{p}$ vs. time, 
corresponding to the 24 choices of payoff matrix
M$^{RSTP}$. The system self-organizes in 
6 different cooperative states with:
$\bar{p}_\infty =1$ (filled lines),
$\bar{p}_\infty \simeq 0.56$ (dotted lines),
$\bar{p}_\infty \simeq 0.5$ (dashed lines), 
$\bar{p}_\infty \simeq 0.42$ (dot-dashed lines), 
$\bar{p}_\infty \simeq 0.22$ (+'s) and
$\bar{p}_\infty \simeq 0.115$ ($\times$'s)
} 
\end{figure}
\end{center}

\vspace{-9mm}

\begin{center}

\begin{tabular}{|p{2,2cm}|p{2cm}|p{2cm}|}
\hline
$R$ $S$ $T$ $P$ & $\bar{p}_\infty$  & $\bar{\delta C}_\infty$  
\\
\hline
 5 3 1 0  $\;$ (C) & 0.425 & 1.87 \\
\hline
 5 3 0 1  $\;$ (C) & 0.113  & 1.15 \\
\hline
 5 1 3 0  $\;$ (C) & 0.42 & 1.81 \\
\hline
 5 1 0 3 & 0.49 & 2.16 \\
\hline
 5 0 3 1 & 0.22 & 1.35\\
\hline
 5 0 1 3 & 0.49 & 2.16  \\
\hline
 3 5 1 0  $\;$ (C) & 1.0 & 3.0 \\
\hline
 3 5 0 1  $\;$ (C) & 1.0 & 3.0  \\
\hline
 3 1 5 0 & 1.0 & 3.0 \\
\hline
 3 1 0 5 & 0.485 & 2.28 \\
\hline
 3 0 5 1  $\;$ (D) & 0.5 & 2.25 \\
\hline
 3 0 1 5 & 0.485 & 2.28 \\
\hline
 1 5 3 0 & 0.51 & 2.22 \\
\hline
 1 5 0 3  $\;$ (C) & 0.5 & 2.25 \\
\hline
 1 3 5 0 & 0.51 & 2.22 \\
\hline
 1 3 0 5 & 0.495 & 2.27 \\
\hline
 1 0 5 3  $\;$ (D) & 0.5 & 2.25 \\
\hline
 1 0 3 5  $\;$ (D) & 0.42 & 2.6 \\
\hline
 0 5 3 1 & 0.5 & 2.22 \\
\hline
 0 5 1 3  $\;$ (D) & 0.5 & 2.23 \\
\hline
 0 3 5 1  $\;$ (D) & 0.51 & 2.22 \\
\hline
 0 3 1 5  $\;$ (D) & 0.495 & 2.25 \\
\hline
 0 1 5 3 & 0.56 & 2.1 \\
\hline
 0 1 3 5 & 0.48  & 2.37  \\
\hline
\end{tabular}

\vspace{1mm}

Table 1. Equilibrium values of probability of cooperation 
$\bar{p}^{R S T P}_\infty$  \&  income-per-agent
$\bar{\delta C}^{R S T P}_\infty$ for the 24 possible 
payoff sets $\{$ {\it R S T P} $\}$. (C) or (D) in first
column indicate if the dominant strategy is C or D.

\end{center}

\vspace{2mm}

{\it The relation between utilities and probability of cooperation} 

\vspace{1mm}

Let us now analyze the average equilibrium income-per-agent 
$\bar{\delta C}_\infty$ for the different payoff matrices. 
The curves of per-capita-income $\delta C^{R S T P}$ as a function of 
the average probability of cooperation $p$ are the parabolas 
obtained by replacing in equation (\ref{eq:Deltacap}) $p_i$ and $p_j$ 
by $p$ {\it i.e.} 
\begin{equation}
\delta C^{R S T P}(p) =  (R-S-T+P) p^2 +(S+T-2P) p +P.
\label{eq:Deltacap2}
\end{equation}
These curves are invariant under the interchange of the
sucker's payoff $S$ and the temptation $T$, {\it i.e.}
\begin{equation}
\delta C^{R S T P}(p) =  \delta C^{R T S P}(p),
\label{eq:symBC}
\end{equation}
{\it i.e.} the 24 payoff matrices give rise to the 12 different parabolas
depicted in Fig. 2. 
In each subplot, the values of $\{ R S T P \}$ and $\{ R T S P \}$
and the equation (\ref{eq:Deltacap2}) for the corresponding parabola
is indicated. 
For example, we have in the first box:
$\delta C^{5310}(p)$ \& $\delta C^{5130}(p) \equiv p^2+4p$.
In all the subplots the equilibrium points 
($\bar{p}_\infty,\bar{\delta C}_\infty$)
for the payoff matrix with $S>T$ and $S<T$ are denoted, respectively, 
by circles and the '+'s.
Note that by virtue that

$$\delta C^{R S T P}(\frac{1}{2})=\frac{R+S+T+P}{4}=\frac{9}{4},$$

all the parabolas $\delta C^{R S T P}(p)$ pass through the point [1/2,9/4].
The values of $\bar{\delta C}^{R S T P}_\infty$ are listed in 
the third column of Table 1 for the 24 payoff matrices.

\begin{center}
\begin{figure}[h]
\centering
\psfig{figure=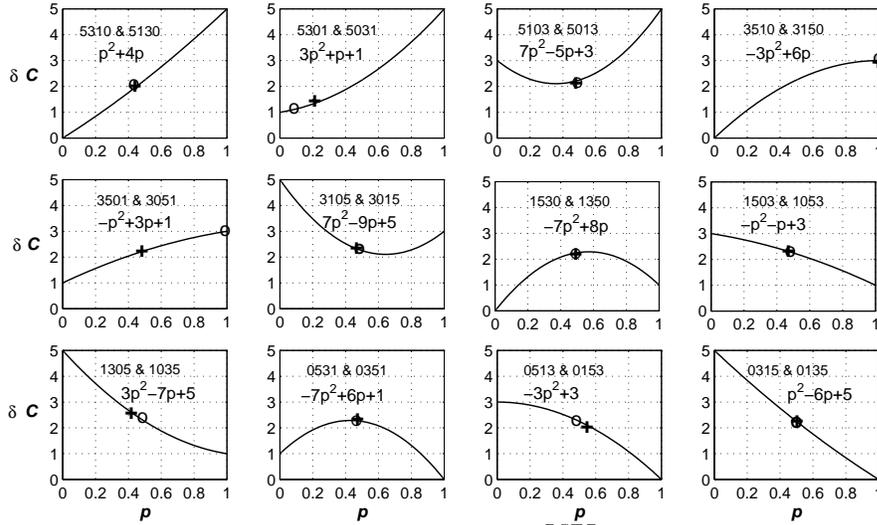,height=7cm}
\caption{The 12 curves of per-capita-income $\delta C^{RSTP}(p)$ 
corresponding to the 24 choices of payoffs R,S,T and P in 
equation (\ref{eq:Deltacap2}).
The payoff matrices M$^{RSTP}$ and M$^{RTSP}$ produce the same parabola. 
In each box is indicated the equation for the parabola and the 
corresponding pair of payoffs which produce it between parenthesis 
($RSTP$ \& $RTSP$).
The quilibrium points ($\bar{p}_\infty,\bar{\delta C}_\infty$), listed in 
Table 1, are marked with circles over the curves 
for the case $S>T$ and with '+'s for the case $S<T$. }
\end{figure}
\end{center}

Let us analyze the distributions of probabilities of cooperation
and their corresponding average capitals and average income-per-agent. 
Fig. 3, Fig.4 and Fig.5 illustrate, respectively, the cases of  
payoff matrices giving rise to equilibrium states with
$\bar{p}_\infty >0.5$, $\bar{p}_\infty\simeq 0.5$ and $\bar{p}_\infty<0.5$.
Measures are performed over 100 
simulations of 50,000 time steps each, after the equilibrium state was 
reached {\it i.e.} typically discarding the first 200,000 
configurations \footnote{The particular case of matrix M$^{3150}$ 
approach to $\bar{p}_\infty=1$ more slowly and after 200,000 iterations 
the system has not reached equilibrium yet as can be seen from Fig.1. In that
case 450,000 iterations were discarded before measuring.}.
The upper plots are distributions for the probabilities of cooperation $p$
using 100-bin histograms. The frequencies $\nu(p)$ are normalized in
such a way that the total area is equal to 1. 
The middle (lower) plots present the corresponding average capitals 
$\bar{C}(p)$ (average income-per-agent 
$\bar{\delta C}(p)$) obtained by taking the quotients between the 
histograms for the capitals ( income-per-agent ) and the frequencies
histograms.  
Fig.3 corresponds to 2 payoff matrices giving rise to high cooperation:
M$^{3150}$ and M$^{0153}$, with $\bar{p}_\infty^{3150}=1$
and $\bar{p}_\infty^{0153} \simeq 0.56$.
The histograms of frequencies $\nu(p)$ exhibit a peak at $p=1$
(in the case of M$^{3150}$
the 3 histograms are non null only for $p=1$).   
Fig. 4 illustrates two cases of moderate cooperation produced by
payoff matrices M$^{3051}$ (the canonical one) and M$^{5310}$, 
with $\bar{p}_\infty^{3051}=0.5$
and $\bar{p}_\infty^{0153}\simeq 0.42$.
Both histograms exhibit two peaks, one at $p=0$ and one at $p=1$.

\begin{center}
\begin{figure}[h]
\centering
\psfig{figure=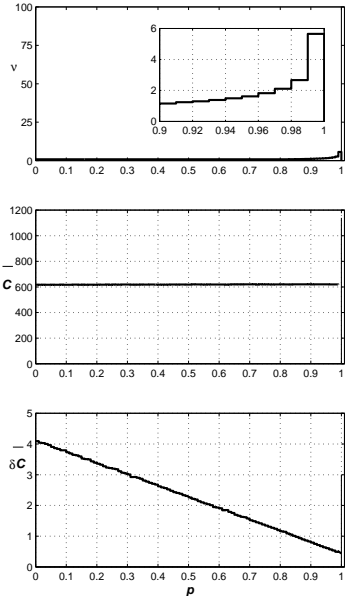,height=8cm}
\caption{100-bin histograms for highly cooperative payoff matrices
 M$^{3150}$ (thin line) and M$^{0153}$ 
(thick line).
Above: Distribution of probabilities of cooperation $p$.  
The inset is a zoom showing the smaller peak at $p=1$ for M$^{0153}$. 
Middle: the corresponding average capitals {\it i.e.} $\bar{C}$ vs. $p$.
Below: the corresponding average income-per-agent {\it i.e.} 
$\bar{\delta C}$ vs. $p$.}
\end{figure}
\end{center}

\begin{center}
\begin{figure}[h]
\centering
\psfig{figure=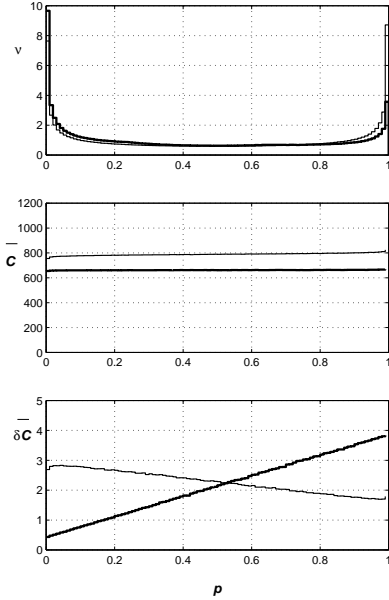,height=8cm}
\caption{100-bin histograms for moderately cooperative payoff matrices
M$^{3051}$ (thin line)  
and M$^{5310}$ (thick line).
Above: Distribution of probabilities of cooperation $p$. 
Middle: the corresponding average capitals {\it i.e.} $\bar{C}$ vs. $p$.
Below: the corresponding average income-per-agent {\it i.e.} 
$\bar{\delta C}$ vs. $p$.}
\end{figure}
\end{center}

\vspace{-5mm}

In Fig.5 are shown two cases of low cooperation, produced by
M$^{5301}$ and M$^{5031}$, with $\bar{p}_\infty^{5301}\simeq 0.113$
and $\bar{p}_\infty^{5031} \simeq 0.22$.
Both histograms of frequencies exhibit a peak at $p=0$.

\begin{center}
\begin{figure}[h]
\centering
\psfig{figure=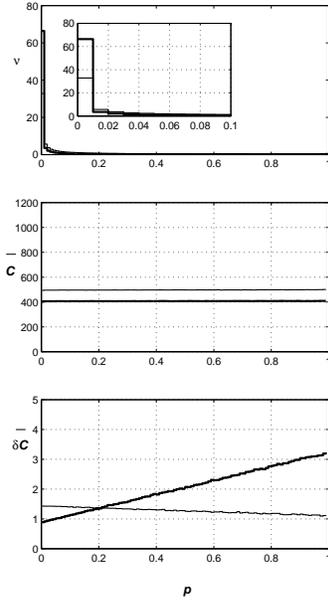,height=8cm}
\caption{100-bin histograms for low cooperative payoff matrices
M$^{5301}$ (thin line) and M$^{5031}$ 
(thick line).
Above: Distribution of probabilities of cooperation $p$.  
The inset is a zoom showing the smaller peak at $p=0$ for M$^{5031}$. 
Middle: the corresponding average capitals {\it i.e.} $\bar{C}$ vs. $p$.
Below: the corresponding average income-per-agent {\it i.e.} 
$\bar{\delta C}$ vs. $p$.}
\end{figure}
\end{center}

\vspace{2mm}

Let us summarize the main results which emerge from the data:

\begin{itemize}
 
\item The state of complete cooperation $\bar{p}_\infty=1$ 
is reached for payoff matrices with $R=3$: 
M$^{3510}$, M$^{3501}$ and M$^{3150}$.
produce the highest possible average cooperation $\bar{p}_\infty=1$.

\item The payoff matrices with the highest possible reward R=5, 
contrary to one might think, do not produce the higher 
$\bar{p}_\infty$. Moreover, the states of lower average cooperation 
is produced by a payoff matrix with $R=5$
: $\bar{p}_\infty=0.115$ occurs 
for M$^{5301}$ and $\bar{p}_\infty=0.22$ for M$^{5031}$. 
The explanation of this fact relies on the adopted measure of success 
based on the estimate $\epsilon(i;t)$: From equation (\ref{eq:LocalEst}) 
and from Fig.2 note that for $R=5$ the estimates, for high values of $p$, 
are $\epsilon > 3$ ({\it i.e.} they are greater than
all the payoffs except the reward), making agents excessively
exigent. In other words: too much rewarding makes the expectation of 
utilities by the agents to be so high that spoils cooperation. 

\item From the two above results it is obvious that there is no
completely clear connection between the dominant strategy and the
equilibrium state. For instance, matrices belonging to class $M_C$
produce both the highest and lowest values of $\bar{p}_\infty$.

\item The highest $\bar{\delta C}_\infty$ is obtained from
payoff matrices which produce the highest $\bar{p}_\infty$,
namely $\delta C^{3 5 1 0}_\infty=
\delta C^{3 5 0 1}_\infty= \delta C^{3 5 0 1}_\infty=3$.
On the other hand, the lowest $\bar{\delta C}_\infty$ is obtained from
payoff matrix which produce the lowest $\bar{p}_\infty$,
namely $\delta C^{5 3 0 1}_\infty\simeq 1.15$.

\item The distributions for the probability of cooperation are clearly non
uniform showing peaks at $p=0$ or/and at $p=1$.

\item The strategy used by the agents is robust enough to lead 
{\em for all} the payoff matrices to $\bar{p}_\infty > 0$. Furthermore, 
for the majority of the 24 payoff matrices $\bar{p}_\infty \gta 0.5$. 
This robustness relies on the strategy combining the 
proposed measure of success and update rule for the probability of
cooperation. Basically it works by tuning the agent's cooperation  
guided by a trade-off between efficiency (increase of utilities) 
and equity (indirect reciprocity). 
If the agent is doing well it behaves nicely and increases its probability 
of cooperation. Nevertheless, in future interactions, if its probability 
of cooperation is inadequate (too high) and it does badly (it is exploited) 
then it reacts by decreasing its cooperation till it starts 
doing well again.

\item The equilibrium states are such that, although 
the average income-per-agent depends on the value of the
probability of cooperation $p$ {\it i.e.} 
$\bar{\delta C}^{R S T P}=\bar{\delta C}^{R S T P}(p)$, 
the distribution of average capitals is almost uniform and does not 
depend on $p$ (as can be observed from
Fig.3 to Fig.5). This is consistent with the fact that agents
constantly adapt their probability of cooperation in such a way to 
improve utilities. Hence, for a given value of $\bar{p}$, the utilities 
of each agent, with probability of cooperation $p$, oscillates around 
$\epsilon(\bar{p})$ in such a way that their accumulated capital 
at a given time (in equilibrium) is independent of $p$.

\end{itemize}

\section{CONCLUSIONS AND FINAL REMARKS}

The first general conclusion is concerning the robustness of the 
cooperative asymptotic state, which
indicates that, in this model,  {\em cooperation seems 
based more in a sort of indirect reciprocity than in selfish incentives}. 
For example, the permutation of the canonical values of $R$ and $T$ 
has the dramatic effect of transforming a society with an intermediate 
level of cooperation into one dominated by defection, as it arises from 
comparing the results for payoff matrices M$^{3051}$ and M$^{5031}$.
On the other hand, the permutation of the canonical values of $S$ 
and $P$ has also a dramatic 
effect: it transforms a society with an intermediate level of 
cooperation into a completely cooperative one, as one can see from 
comparing the results for payoff matrices M$^{3051}$ and M$^{3150}$.

An interesting extension of the model would be to allow competition
of different strategies to promote their evolution {\it i.e.} players 
which imitate the best-performing ones in such a way that lower 
scoring strategies decrease in number and the higher scoring increase.
In particular, a possibility would be to associate 
different strategies with the use of disctint payoff matrices.
For instance, individuals inclined to cooperate (defect) might 
be represented by agents using the payoff matrix M$^{3150}$ (M$^{5301}$)
while "neutral" agents by agents using the canonical payoff 
matrix M$^{3051}$. 
This would make possible to study 
if mutants inclined to D can invade a group of neutral individuals or
individuals inclined to C and drive out all cooperation.
However, a previous necessary step was the knowledge of the effect of 
changing the payoff matrix on the system self-organization and in 
particular on the equilibrium point 
($\bar{p}^{RSTP}_\infty,\bar{\delta C}^{RSTP}_\infty$). 
So in this work we considered each of these 24 payoff matrices by separate.

This model can be extended in other ways in order to make it
more realistic. 
For instance, here I considered a MF approximation which   
neglects all the spatial correlations. One virtue of this 
simplification is that it shows the model does not require
that agents interact only with those within some geographical proximity 
in order to sustain cooperation. Playing with fixed neighbors
is sometimes considered as an important ingredient to
successfully maintain the cooperative regime \cite{cra2001},\cite{nm92}.
However, the quality of this MF approximation
depends on the nature of the system one desires to model
(people, cultures of bacteria, market of providers of different services
or products, etc.).
Therefore, in order to apply the model to situations in which the
effect of geographic closeness cannot be neglected 
an interesting extensions of the model would be: 
to transform the entirely random  PD game into a spatial PD game, 
in which individuals interact only (or mainly) with those within some 
geographical proximity. 

To conclude, this work is based on the canonical assumption that 
individuals are entirely self-interested. However, recent investigations, 
performed in twelve countries on four continents, have uncovered 
systematic deviations from the material payoff-maximizing
{\it dogma}\cite{hbbcfgm01}. In addition to their own material payoffs, 
many experimental subjects appear to prefer to share resources and
undertake costly reciprocal actions in anonymous one-shot interactions. 
Therefore, an open issue is how to incorporate this fact in 
a more realistic model.

\vspace{2mm}

{\bf Acknowledgments.}

I thank R. Axelrod and R. Donangelo for valuable discussions and opinions.
I thank M. Reisenberger for criticism of the manuscript.

\end{document}